
\magnification 1200
\def \v {\vert}
\def \ra {\rangle}
\def \a {\alpha}
\def \b {\beta}
\baselineskip=24pt
\line{\hfill Nov 1994}
\vskip 3 true cm
\centerline{\bf An optical method for teleportation}
\vskip 3 true cm
\centerline{Sandu Popescu}
\vskip 2 true cm
\centerline{
School of Physics and Astronomy,
Tel Aviv University,
Tel Aviv, Israel}
\centerline{and}
\centerline{
Department of Physics, Boston University, Boston, MA 02215.}
\vskip 4 true cm
\noindent
A teleportation method using standard present day optical technology
is presented.
\vfill
\eject
About two years ago, C. Bennett and collaborators
invented ``teleportation" [1], a method for transmitting quantum
states from
one place to another by using a ``nonlocal" communication channel in
parallel with a standard classical one.
Ever since its invention there was a strong desire for the experimental
verification of teleportation, both in order to test such a fundamental
aspect of quantum mechanics (the nonlocality revealed by teleportation
is different from the one revealed by the usual Bell type measurements
[2]) and also in view of its potential
applications in quantum information transmission, quantum cryptography, etc.
But the experimental realization of teleportation is not easy. The main
difficulty lies in implementing the transmitter's actions. According to the
original scheme, the transmitter has to measure a quantum
operator A which acts on two different particles  and
whose (nondegenerate) eigenstates are entangled (non direct-product)
states of these two particles. A similar problem arises also in the
alternative teleportation method proposed by L. Vaidman [3]. But despite the
recent progress in
manipulating entangled states - is is now relatively easy to prepare two
particles in a given entangled state (the most common case being that of
two photons) - no one has measured yet such an operator.

At present there are a few proposal for experimental verification of
teleportation. Some of them rely on some very new and rather exotical
techniques such as atoms interacting with electro-magnetic cavities [4-6]
and optical parametric up-conversion (the opposite of the already common
technique of optical parametric down conversion) [7]. These schemes are
supposed to implement the original teleportation scheme as described in
[1]. There exist also experiment proposals [8,9] based on more
traditional quantum optical techniques, but these schemes differ from
the original one and, even in the ideal limit, only 50\% of the attempted
teleportations succeed while in the other cases the state which has to
be teleported is completely and irremediably destroyed.

In the present paper I propose yet another teleportation scheme
based on standard quantum optical technology but which achieves,
in the ideal limit, 100\% success. This scheme is different from the
original one in the way in which the state which has to be transmitted
is presented to the transmitter.  The original teleportation scheme
involves three observers, a Preparer, a Transmitter and a Receiver (the
last two usually called Alice and Bob), and three particles: a pair of
particles in an entangled state (shared by Alice and Bob and constituing
the ``nonlocal transmission channel"), and a
particle originally prepared by the Preparer in a state $\Psi$ and then
given to Alice. (The state $\Psi$ is unknown for to Alice). In the
present scheme
instead of encoding $\Psi$ in a third particle, the
Preparer encodes the state $\Psi$ in a different degree of freedom of
Alice's member of the pair shared by her and Bob. This avoids altogether
the difficult
problem originally facing Alice, that is, the measurement of an operator
acting on two particles and having entangled eigenstates. Instead Alice
has to measure a formally identical operator but acting this time on two
different degrees of freedom of the same particle - a much simpler task.
Following I shall first describe the experimental proposal and after
that I will discuss the significance of my departure from the original
teleportation scheme.

The optical setup is illustrated in fig. 1. The first stage is to
produce by standard parametric down-conversion two photons both with the
same polarization, say horizontal (h), and entangled in directions, such
that their initial state is
$$\v\Phi_{in}\ra={1\over{\sqrt2}}\bigl(\v a\ra_1\v a'\ra_2+\v b\ra_1\v
b'\ra_2\bigr)\v{\bf h}\ra_1\v{\bf h}\ra_2,\eqno(1)$$
where $\v a\ra_1$ and $\v b\ra_1$ represent photon 1 being in beams $a$
and $b$ respectively, and similarly $\v a'\ra_2$ and $\v b'\ra_2$ represent
photon 2 in beams $a'$ and $b'$ respectively. The boldfaced vectors represent
the polarization, in this case horizontal, of photons 1 and 2. One of
these photons, say photon 1, is sent to Alice and the other one to Bob.
These two photons represent the ``nonlocal transmission channel", the
equivalent of the two entangled spin 1/2 particles in the original
teleportation paper.

On the way to Alice, photon 1 is intercepted by the Preparer who changes
its polarization from horizontal to some arbitrary (not necessary linear)
polarization $\v {\bf \Psi}\ra_1$. The Preparer affects the polarization
in both beams a and b in the same way.
The state ${\bf \Psi}$ is the state Alice has to transmit to Bob.

After having the polarization changed, photon 1 reaches Alice. The state
of the two photons, just before reaching Alice is thus
$$\v \Phi\ra={1\over{\sqrt2}}\bigl(\v a\ra_1\v a'\ra_2+\v b\ra_1\v
b'\ra_2\bigr)\v{\bf \Psi}\ra_1\v{\bf h}\ra_2,\eqno(2)$$
which is the formal analogue of state $\Psi_{123}$ in [1] (eq. 4 in [1]).
The polarization degree of freedom in which the state given by the
Preparer is encoded is totally uncorrelated to the direction degrees of
freedom in which the two photons are entangled.

If photon storage and photon-photon interaction would be an easy thing
to do, the Preparer would not have had to intercept the photon Alice
will use for transmission but could have given her another photon, say
photon 0 with polarization ${\bf \Psi}$. Alice could have then
stored this photon and after receiving photon1 she could have
interchanged their polarization
$$\v{\bf \Psi}\ra_0\v{\bf h}\ra_1\rightarrow\v{\bf h}\ra_0\v{\bf
\Psi}\ra_1\eqno(3)$$
and then proceed with the method I will describe bellow. In such a case
teleportation would have been realized exactly like originally
proposed. Unfortunately photon storage and photon-photon interactions
are very difficult to realize in practice, and the Preparer has to help
Alice by encoding the unknown state ${\bf \Psi}$ directly in the
polarization of photon 1.

Now, following the teleportation procedure, Alice has to measure the
operator $A$ whose (nondegenerate) eigenstates are

$${1\over{\sqrt2}}\bigl(\v a\ra_1\v {\bf v}\ra_1+\v b\ra_1\v {\bf
h}\ra_1\bigr)$$

$${1\over{\sqrt2}}\bigl(\v a\ra_1\v {\bf v}\ra_1-\v b\ra_1\v {\bf
h}\ra_1\bigr)$$

$${1\over{\sqrt2}}\bigl(\v a\ra_1\v {\bf h}\ra_1+\v b\ra_1\v {\bf
v}\ra_1\bigr)$$

$${1\over{\sqrt2}}\bigl(\v a\ra_1\v {\bf h}\ra_1-\v b\ra_1\v {\bf
v}\ra_1\bigr). \eqno(4)$$

The measurement of the operator $A$ effectively implies an interaction
between the two degrees of freedom (direction and polarization) of
photon 1. Indeed, suppose that we start with the state $\v a\ra_1\v {\bf
v}\ra_1$, measure $A$ and obtain, say, $A=a_1$, the eigenvalue
corresponding to the first
eigenstate (4). Then after the measurement photon 1 is in state
${1\over{\sqrt2}}\bigl(\v a\ra_1\v {\bf v}\ra_1+\v b\ra_1\v {\bf
h}\ra_1\bigr)$, meaning that the direction and the polarization got
entangled.  In the original teleportation paper the operator $A$ acts on
two different particles and effectively implies an interaction between
them. But controlled interaction between two quantum particles is much
more difficult to realize - that's why the experimental implementation
of the original teleportation method is so difficult. On the other hand,
to realize an interaction between the direction and the polarization of
the same photon is trivially simple: any polarizing beam splitter does
that.

To measure $A$ Alice first splits each of her incoming beams, $a$ and
$b$, into two beams, by use of polarizing beam-splitters and then, once
the information about polarization is encoded in the position of the
photon, she rotates the polarization in all the beams to the same
direction. These unitary transformations, as illustrated in fig.1, are

$$\v a\ra_1\v{\bf v}\ra_1\rightarrow
\v 1\ra_1\v{\bf v}\ra_1\rightarrow\v 1\ra_1\v{\bf h}\ra_1$$
$$\v a\ra_1\v{\bf h}\ra_1\rightarrow
\v 2\ra_1\v{\bf h}\ra_1\rightarrow\v 2\ra_1\v{\bf h}\ra_1$$
$$\v b\ra_1\v{\bf v}\ra_1\rightarrow
\v 3\ra_1\v{\bf v}\ra_1\rightarrow\v 3\ra_1\v{\bf h}\ra_1$$
$$\v b\ra_1\v{\bf h}\ra_1\rightarrow
\v 4\ra_1\v{\bf h}\ra_1\rightarrow\v 4\ra_1\v{\bf h}\ra_1. \eqno(5)$$
Measuring the operator $A$ reduces now to measuring an operator $A'$ acting
solely on the position of photon 1 and having the eigenstates
$${1\over{\sqrt2}}\bigl(\v 1\ra_1+\v 4\ra_1\bigr)$$
$${1\over{\sqrt2}}\bigl(\v 1\ra_1-\v 4\ra_1\bigr)$$
$${1\over{\sqrt2}}\bigl(\v 2\ra_1+\v 3\ra_1\bigr)$$
$${1\over{\sqrt2}}\bigl(\v 2\ra_1-\v 3\ra_1\bigr).\eqno(6)$$
This can easily be done by letting beams 1 and 4 impinge onto a symmetric
beam-splitter $S_1$, followed by detectors $D_1$ and $D_4$, and beams 2 and
3 impinge onto beam-splitter $S_2$ followed by detectors $D_2$ and $D_3$.
Indeed, the symmetric beam-splitters $S_1$ and $S_2$ transform the input
states $\v 1\ra_1$ and $\v 4\ra_1$ and respectively $\v 2\ra_1$ and $\v
3\ra_1$ into the corresponding linear transformations
$${1\over{\sqrt2}}\bigl(\v 1\ra_1+\v 4\ra_1\bigr)\rightarrow\v 1'\ra_1$$
$${1\over{\sqrt2}}\bigl(\v 1\ra_1-\v 4\ra_1\bigr)\rightarrow\v 4'\ra_1$$
$${1\over{\sqrt2}}\bigl(\v 2\ra_1+\v 3\ra_1\bigr)\rightarrow\v 2'\ra_1$$
$${1\over{\sqrt2}}\bigl(\v 2\ra_1-\v 3\ra_1\bigr)\rightarrow\v
3'\ra_1\eqno(7)$$
which are directly detected by $D_1 - D_4$.

To summarize, if the polarization state prepared by the Preparer was
$$\v {\bf \Psi}\ra_1=\a\v {\bf h}\ra_1+\b\v{\bf
v}\ra_1,\eqno(8)$$
then after all the actions made by Alice, the state of the two photons
just before the detection of photon 1 by one of the detectors $D_1 -
D_4$ is
$${1\over2}\bigl\{\v 1'\ra_1(\b\v a'\ra_2+\a\v b'\ra_2)+
\v 2'\ra_1(\a\v a'\ra_2+\b\v b'\ra_2)+
\v 3'\ra_1(\a\v a'\ra_2-\b\v b'\ra_2)+$$
$$\v 4'\ra_1(\b\v a'\ra_2-\a\v b'\ra_2)\bigr\}\v{\bf h}\ra_1\v{\bf
h}\ra_2,\eqno(9)$$
the analog of Eq.(5) in the original paper [1].
Therefore,  depending on which detector registers photon 1,  photon 2 is
left in one of the four states
$$\bigl(\b\v a'\ra_2+\a\v b'\ra_2\bigr)\v{\bf h}\ra_2,$$
$$\bigl(\a\v a'\ra_2+\b\v b'\ra_2\bigr)\v{\bf h}\ra_2,$$
$$\bigl(\a\v a'\ra_2-\b\v b'\ra_2\bigr)\v{\bf h}\ra_2,$$
$$\bigl(\b\v a'\ra_2-\a\v b'\ra_2\bigr)\v{\bf h}\ra_2 \eqno(10)$$
which are isomorphic to ${\bf\Psi}$ up to some standard unitary
transformations, independent of ${\bf\Psi}$.

Having finished with Alice's actions, let us now analyze Bob's.
Bob's task is to reconstruct the polarization state ${\bf \Psi}$ into
his photon.
To this end he first translates the information which is contained in
the states (10) in the direction degree of freedom into polarization.
This is done by rotating the polarization in beam $a'$ from horizontal to
vertical, and then merging the two beams $a'$ and $b'$ into a single one
(denoted by o) by use of a polarized beam splitter.
That is, Bob's actions are

$$\v a'\ra_2\v{\bf h}\ra_2\rightarrow\v a'\ra_2\v{\bf v}\ra_2\rightarrow
\v o\ra_2\v{\bf v}\ra_2$$
$$\v b'\ra_2\v{\bf h}\ra_2\rightarrow\v b'\ra_2\v{\bf h}\ra_2\rightarrow
\v o\ra_2\v{\bf h}\ra_2.\eqno(11)$$
As a result the four possible states (10) are transformed into
$$\bigl(\b\v a\ra_2+\a\v b\ra_2\bigr)\v{\bf h}\ra_2
\rightarrow\v o \ra_2\bigl(\b\v{\bf v}\ra_2+\a\v{\bf h}\ra_2\bigr)
$$
$$\bigl(\a\v a\ra_2+\b\v b\ra_2\bigr)\v{\bf h}\ra_2
\rightarrow\v o \ra_2\bigl(\a\v{\bf v}\ra_2+\b\v{\bf h}\ra_2\bigr)$$
$$\bigl(\a\v a\ra_2-\b\v b\ra_2\bigr)\v{\bf h}\ra_2
\rightarrow\v o \ra_2\bigl(\a\v{\bf v}\ra_2-\b\v{\bf h}\ra_2\bigr)$$
$$\bigl(\beta\v a\ra_2-\alpha\v b\ra_2\bigr)\v{\bf h}\ra_2
\rightarrow\v o \ra_2\bigl(\beta\v{\bf v}\ra_2-\alpha\v{\bf h}\ra_2\bigr).
\eqno(12)$$
Photon 2 is then kept in flight, by sending it on a long optical
path (illustrated by the dotted line in fig. 1) until the signal comes
from Alice which of the detectors $D_1 - D_4$ has registered photon 1.
This signal is processed by some  fast electronic device which
can then activate some appropriate active optical elements so that when
photon 2 goes through them its polarization is rotated
to $\v{\bf \Psi}\ra_2$. This could be obtained for
example by two active optical elements (such as two Kerr or two Pockel
cells) placed one after the other (C1 and C2 on fig. 1) where the first
one just reverses the signs of the horizontal polarization
$$\v {\bf v}\ra_2\rightarrow\v {\bf v}\ra_2$$
$$\v {\bf h}\ra_2\rightarrow-\v {\bf h}\ra_2 \eqno(13)$$
while the other rotates horizontal into vertical polarization and
vice-versa,
$$\v {\bf v}\ra_2\rightarrow\v {\bf h}\ra_2$$
$$\v {\bf h}\ra_2\rightarrow\v {\bf v}\ra_2. \eqno(14)$$
By activating none, one, or both of these cells, according to the result
obtained by Alice, Bob accomplishes his task.

The success of teleportation can be verified by letting Bob's photon
impinge on a polarizer parallel with the one used by the Preparer -
Bob's photon must pass it with certainty. In fact even if one does not
succeed to implement the rapid communication between Alice and Bob, and
the on-flight rotation of Bob's photon, one could still verify the
success of the non-local part of the transmission. In such a case Bob's
station could end with the photon passing through the polarizing
beam-splitter (SP3) and from here the photon enters the verification
station. The verification station consists simply of a polarizer which
is set at random in one of the four positions which could transmit
with certainty one of the states (12). (The four states (12) are not
orthogonal on each other, that's why one cannot devise a measurement
which can tell them apart. The most one can do is to set a polarizer
such as one of these states passes with certainty, while the others have
some smaller probability to pass.)  To verify the success of
teleportation one should analyze the subensemble of runs in which it
happened that Alice got one of the four results corresponding to the
particular setting of the verification polarizer in that run. In this
subensemble Bob's photon must pass the verification polarizer with
certainty. (In fact if we just wish to check the success of the
non-local part of the teleportation and do not request the actual
reconstruction of the polarization state ${\bf \Psi}$, there is no need
for Bob's station at all. One could let Bob's input beams $a'$ and $b'$ to
enter directly in the verification station, which consists of an
arrangement of beam-splitters, phase shifters and detectors which could
check, at random, for one of the four states (10). Then, as in the
previous case, one should verify that in the subensemble of cases in
which it so happened that Alice got the result corresponding to the
particular state which was verified, Bob's photon has passed the
verification with certainty.

At first sight the teleportation method presented in this paper seems
very different from the original scheme. In the original method the
Preparer need not know any details about the particular set-up which
will be used for teleportation.  He just has to prepare his own particle
in the state he wants to have transmitted and then give it to Alice. On
the other hand, in the above presented method the Preparer has to encode
the state which has to be transmitted in the very particle which will be
used for transmission. But actually the difference between the two
methods is limited only to a local pre-processing of data at the level
of the transmitter (Alice), and has nothing to do with the process of
transmission itself, i.e.  with the teleportation proper. As in the
original method Alice is faced with the same problem - she has only a
single replica of the state ${\bf \Psi}$ and thus she cannot find out
what this state is - and she solves it in an identical way - she doesn't
even try to find out ${\bf \Psi}$ but she sends it directly to Bob by
using a ``nonlocal communication channel" in parallel with a
conventional one. So the above presented method deserves the name of
teleportation.

In conclusion, I have presented a method for teleportation which can be
implemented with present day optical technology. The basic idea is to
have the Preparer helping Alice by encoding the state which has to be
transmitted directly in Alice's member of the entangled photons which
will be used for transmission. Actually this idea is more important than
the particular experimental set-up I have described in the present
paper. Using the same idea many other set-ups can be designed. An
optimization of the above method has already been proposed by H.
Bernstein [10]. As yet another possibility, instead of having the two
photons entangled in directions and the state $\Psi$ encoded in the
polarization degree of freedom, the photons could be entangled in
polarization and then $\Psi$ encoded in the direction of one of them (by
having the Preparer use a beam-splitter and phase shifters).

Finally, the method presented in this paper modified such that Alice
plays also the role of the Preparer and Bob
performs randomly chosen measurements on his photon outside the light-cone
of Alice's measurements can be used as a generalized Bell's inequalities
type  measurement. This is a generalized Bell type measurement because
Alice performs a generalized measurement (a
POVM [11]) on her photon, instead of simply measuring operators
which view her photon as a 2-dimensional system (i.e. taking into
account only the fact that her photon lives in beams a or
b) as in a usual Bell type measurement carried on a pair of photons
emerging from a parametric down-conversion source. It is very possible
that such a generalized Bell type measurement could solve the detector
efficiency problem which does not allow in a usual Bell type measurement
to reach a conclusion in favor of quantum mechanics or local hidden
variables models with present day detectors - as Alice can choose among
more measurements than in a usual Bell measurement, it might be that
contradictions between quantum mechanics and local hidden variables
models could be already be observed even with low detector efficiencies.

\bigskip
\noindent
Acknowledgment. I would like to acknowledge the support of NSF Grant
PHY-9321992.
\bigskip
\noindent
References.
\bigskip
\noindent
1. C. Bennett, G. Brassard, C. Crepeau, R. Jozsa, A. Peres and W.
Wootters, Phys. Rev. Lett. 70, 1895, (1993).
\bigskip
\noindent
2. S. Popescu, Phys. Rev. Lett. 72, 797, (1994).
\bigskip
\noindent
3. L. Vaidman, Phys. Rev. A 49, 1473 (1994).
\bigskip
\noindent
4. L. Davidovich, N. Zagury, M. Brune, J. M. Raimond and S. Haroche,
Phys. Rev. A, 50, 895, (1994).
\bigskip
\noindent
5. T. Sleator and H. Weinfurter, in IQEC Technical Digest 1994, Vol. 9,
1994, OSA Technical Digest Series (OSA, Washington, D.C., 1994), p. 140.
\bigskip
\noindent
6. J. I. Cirac and A. S. Parkins, Lecture presented at the Workshop on
Quantum Computing and Communication, Gaithersburg, MD, August 18-19, 1994.
\bigskip
\noindent
7. C. H. Bennett, private communication.
\bigskip
\noindent
8. A. Mann and S. L. Braunstein, Measurement of the Bell operator and
quantum teleportation, Preprint, Technion - Israel Institute
of Technology, Haifa, Israel.
\bigskip
\noindent
9. H. Weinfurter, private communication.
\bigskip
\noindent
10. H. Bernstein, private communication.
\bigskip
\noindent
11. J.M. Jauch and C. Piron, Helvetica Phys. Acta 40 559 (1967); C.W.
Helstrom, Quantum Detection and Estimation Theory, Academic Press, N.Y.
1976, pp 74-83; A. Peres, Quantum Theory: Concepts and Methods, Kluwer,
Dordrecht, 1993, pp 282-289.
\end